# Breyer case of the Court of Justice of the European Union: IP addresses and the personal data definition

Frederik Zuiderveen Borgesius[1]

*Court of Justice of the European Union, Case C-582/14, Patrick Breyer v. Bundesrepublik Deutschland, judgment of 19 October 2016, not yet published. ECLI:EU:C:2016:779*

*A dynamic IP address of a website visitor is a piece of personal data for a website publisher, if the publisher has the legal means to identify the visitor, with the help of additional information held by the visitor's internet access provider.*

*Article 2(h) and article 7(f) of the Data Protection Directive*

The Breyer case of the Court of Justice of the European Union (CJEU) primarily concerns the question whether a website visitor's dynamic IP address constitutes personal data for a website publisher, when another party (an internet access provider) can tie a name to that IP address. In essence, the Court finds that an IP address constitutes personal data for the website publisher, if that publisher has the legal means to obtain, from the visitor's internet access provider, additional information that enables the publisher to identify that visitor. Below I summarise the facts and the judgment, and add a few comments.

---

[1] Dr. Frederik Zuiderveen Borgesius professor ICT and law and iHub, Radboud University. Frederik[at]cs.ru.nl. Frederik would like to thank Damian Clifford, João Pedro Quintais, and Tijmen Wisman for their helpful comments.

## 1.1 Facts

Mr. Breyer is an activist and politician.[2] He visited several websites of German Federal institutions. Most of those websites store logfiles regarding website visitors. These logfiles include 'the name of the web page or file to which access was sought, the terms entered in the search fields, the time of access, the quantity of data transferred, an indication of whether access was successful, and the IP address of the computer from which access was sought.'[3] The website publishers store these logfiles to prevent denial-of-service attacks, and, sometimes, to bring criminal proceedings against attackers.

IP addresses are, in the words of the CJEU, 'series of digits assigned to networked computers to facilitate their communication over the internet.'[4] IP addresses can be 'static' or 'dynamic'. Static IP addresses, also called fixed IP addresses, are 'invariable and allow continuous identification of the device connected to the network'.[5]

By contrast, a dynamic IP address 'changes each time there is a new connection to the internet'.[6] As the Advocate General notes, internet access providers 'keep a record of which IP address has been assigned, at any one time, to a particular device'.[7] The CJEU adds: 'Unlike static IP addresses, dynamic IP addresses do not enable a link to be established, through files accessible to the public, between a given computer and the physical connection to the network used by the internet service provider.'[8]

---

[2] Unrelated to this case, Breyer (of the Piratenpartei) published a paper on surveillance in 2005: Breyer P, 'Telecommunications data retention and human rights: the compatibility of blanket traffic data retention with the ECHR' (2005) 11(3) European Law Journal 365 <http://www.daten-speicherung.de/data/data_retention_and_human_rights_essay.pdf>.
[3] Para 14 of the judgment.
[4] Para 15.
[5] Para 36.
[6] Para 16.
[7] Para 2 of the Advocate General's opinion.

## 1.2   Procedure in Germany

Breyer went to court in Germany. In short, he wanted the court to stop the Federal Republic of Germany from storing his IP address, except when the storing is necessary to restore the availability of a website in the event of a fault.[9] Breyer lost the case in first instance. He appealed.

The court of appeal found that the storing of Breyer's IP address did not render him identifiable for the website publisher (in this case the Federal Republic of Germany) and therefore did not come within the definition of personal data.[10]

But the court of appeal also partly ruled in favour of Breyer. Sometimes website visitors provide their name (or other directly identifying information) to a website publisher, for instance when they subscribe to a newsletter. In such situations, the website publisher can tie the name to the website visitor's IP address: the visitor would be identifiable for the website publisher.[11]

For such situations, the court of appeal ordered the website publisher to refrain from storing, at the end of each consultation period, Breyer's IP address, except when that storage is necessary to restore the dissemination of website data in the event of a fault occurring.[12]

---

[8] Para 16 of the judgment.
[9] It appears that the English version of the judgment contains two typos (the version that was online on 18 February 2017). In para 17, the phrase "except in so far as its storage is unnecessary in order to restore the availability of those media in the event of a fault occurring" should be: "except in so far as its storage is *necessary* in order to restore the availability of those media in the event of a fault occurring". In para 19, the phrase "except in so far as that storage is not necessary in order to restore the dissemination of those media in the event of a fault occurring" should be: "except in so far as that storage is *necessary* in order to restore the dissemination of those media in the event of a fault occurring". As the original language of the case, German is the authentic language of the judgment (see Article 41 of the Rules of Procedure of the Court of Justice of the European Union).
[10] Para 21 of the jugdment.
[11] Para 20.
[12] Para 19.

Both Breyer and the Federal Republic of Germany (the website publisher) appealed to the Bundesgerichtshof (Federal Court of Justice, Germany). Breyer wanted to have his application for an injunction upheld in its entirety; the Federal Republic of Germany wanted to have it dismissed.

## 2    Bundesgerichtshof on IP addresses

The Bundesgerichtshof finds that Breyer's IP address does not enable the website publisher to *directly* identify Breyer. According to the Bundesgerichtshof, the website publishers can identify Breyer only if the information relating to his identity is communicated to them by his internet access provider. The classification of the IP address as 'personal data' thus depends on whether Breyer is *identifiable*.

The Bundesgerichtshof refers to legal scholarship in Germany that distinguishes an 'objective' and a 'relative' criterion to assess whether IP addresses should be seen as personal data. Academics disagree on which criterion should be used.

Using an objective criterion, 'the IP addresses at issue in the main proceedings may be regarded, at the end of the period of use of the websites at issue, as being personal data even if only a third party is able to determine the identity of the data subject, that third party being, in the present case, Mr Breyer's internet service provider, which stored the additional data enabling his identification by means of those IP addresses.'[13]

However, using a relative approach, the IP address would only be a piece of personal data for Breyer's internet access provider, and not for the website publisher. In the words of the CJEU: 'According to a "relative" criterion, such data may be regarded as personal data in relation to an entity such as Mr Breyer's internet service provider because they allow the user to be precisely identified (…), but not being regarded as such with respect to another entity, since that operator does not have, if Mr Breyer has not disclosed his identity during the consultation of those websites, the information necessary to identify him without disproportionate effort.'[14]

---

[13] Para 25.

For guidance on interpreting the personal data definition, the Bundesgerichtshof referred the following question (the first of two) to the CJEU:

> 'Must Article 2(a) of [the Data Protection Directive[15]] be interpreted as meaning that an internet protocol address (IP address) which an [online media] service provider stores when his website is accessed already constitutes personal data for the service provider if a third party (an access provider) has the additional knowledge required in order to identify the data subject?'[16]

## 3    CJEU on IP addresses

First, the CJEU turns to the Data Protection Directive's personal data definition:

> 'Personal data' shall mean any information relating to an identified or identifiable natural person ('data subject'); an identifiable person is one who can be identified, directly or indirectly, in particular by reference to an identification number or to one or more factors specific to his physical, physiological, mental, economic, cultural or social identity.[17]

---

[14] Para 25.
[15] Amended by the author. Data Protection Directive 95/46, O.J.1995, L281.
[16] Para 30. The CJEU rephrases the question, as it often does, and limits the question to dynamic IP addresses: 'the referring court asks essentially whether Article 2(a) of [the Data Protection Directive] must be interpreted as meaning that a dynamic IP address registered by an online media services provider [i.e. a website publisher] when a person accesses a website that that provider makes accessible to the public constitutes, with regard to that service provider, personal data within the meaning of that provision, where, only a third party, in the present case the internet service provider, has the additional data necessary to identify him '(para 31).
[17] Article 2(a) of the Data Protection Directive.

In Scarlet/Sabam (2011), the CJEU said about IP addresses in the hands of an internet access provider that those addresses were personal data. An internet access provider can usually tie the name of its customer to the IP address that the provider has allocated to the customer.[18] In the Breyer case, however, the question is whether a website publisher (rather than an internet access provider) can identify Breyer.[19]

The CJEU summarises that the Bundesgerichtshof's question is based on two assumptions. The question is based 'on the premise, first, that data consisting in a dynamic IP address and the date and time that a website was accessed from that IP address registered by an online media services provider do not, without more, give the service provider the possibility to identify the user who consulted that website during that period of use'.[20] Second, the Bundesgerichtshof assumes that Breyer's internet access provider has additional information that could be combined with the IP address to identify the website visitor.[21] The CJEU adds that 'it is common ground that the IP addresses to which the national court refers are 'dynamic' IP addresses'.[22]

Under those assumptions, Breyer's dynamic IP address is not information relating to an 'identified' person for the website publisher, says the CJEU. After all, the IP address does 'not directly reveal the identity of the natural person who owns the computer from which a website was accessed, or that of another person who might use that computer.'[23]

Therefore, the CJEU observes that the question in fact refers to whether a dynamic IP address, stored by a website publisher, 'may be treated as data relating to an "identifiable natural person" where the additional data necessary in order to identify the user of a website that the [publisher] makes accessible to the public are held by that user's internet service provider.'[24]

---

[18] See opinion Advocate General, para 2.
[19] CJEU, judgment of 24 November 2011, *Scarlet Extended* (C-70/10, EU:C:2011:7, para 51.
[20] Para 37 of the judgment. The original quotation uses "premiss", rather than "premise".
[21] Para 37.
[22] Para 36.
[23] Para 38.
[24] Para 39.

The CJEU notes that, under the Data Protection Directive's personal data definition, 'an identifiable person is one who can be identified, directly or indirectly'.[25] The CJEU adds: 'The use by the EU legislature of the word "indirectly" suggests that, in order to treat information as personal data, it is not necessary that that information alone allows the data subject to be identified.'[26]

Then the CJEU turns to recital 26 of the Directive, which reads as follows: 'to determine whether a person is identifiable, account should be taken of all the means likely reasonably to be used either by the controller *or by any other person to identify the said person* (…)'.[27] The CJEU concludes: 'for information to be treated as "personal data" (…), it is not required that all the information enabling the identification of the data subject must be in the hands of one person.'[28]

Recital 26 also states: 'to determine whether a person is identifiable, *account should be taken of all the means likely reasonably to be used* either by the controller or by any other person to identify the said person'.[29] Therefore, the CJEU assesses next 'whether the possibility to combine a dynamic IP address with the additional data held by the internet [access] provider constitutes a means likely reasonably to be used to identify the data subject.'[30]

According to the CJEU, the criterion of the '*means likely reasonably to be used*' would not be satisfied 'if the identification of the data subject was prohibited by law or practically impossible on account of the fact that it requires a disproportionate effort in terms of time, cost and man-power, so that the risk of identification appears in reality to be insignificant.'[31]

The CJEU notes that, in the event of denial-of-service attacks, website publishers have legal channels in Germany to obtain identifying information from internet access providers, to identify website users and to bring criminal proceedings.[32] Therefore it appears that the website publisher 'has the means which may likely reasonably be used in order to identify the data subject, with the assistance of other persons, namely the competent authority and the internet service provider, on the basis of the IP addresses stored.'[33]

---

[25] Para 40; article 2(a) of the Data Protection Directive.
[26] Para 41.
[27] Recital 26 of the Data Protection Directive (emphasis added); para 42 of the judgment.
[28] Para 43.
[29] Emphasis added.
[30] Para 45.
[31] Para 46. The CJEU refers to para 46 of the AG's opinion

The CJEU concludes by answering the Bundesgerichtshof's first question as follows:

> 'a dynamic IP address registered by an online media services provider [i.e. a website publisher] when a person accesses a website that the provider makes accessible to the public constitutes personal data within the meaning of that provision, in relation to that provider, where the latter has the legal means which enable it to identify the data subject with additional data which the internet service provider has about that person.'[34]

## 4    Bundesgerichtshof on the balancing provision (article 7(f) of the Data Protection Directive)

The second question referred to the CJEU by the Bundesgerichtshof concerns the interpretation of article 7(f) of the Data Protection Directive, also called the balancing provision.

Germany has strict rules on storing IP addresses. Roughly summarised, the German Telemediengesetz (Law on telemedia) only allows website publishers to store IP addresses if the website visitor has consented to that storage, or when a specified

---

[32] Para 47.
[33] Para 48.
[34] Para 49 and dictum; amendment between square brackets by the author.

exception in the Telemediengesetz applies: for billing, or for ensuring that the website works well.[35]

The Data Protection Directive lays down another regime. The Directive allows personal data processing if the processing complies with all the Directives requirements. One of those requirements is that personal data processing must be based on a legal basis. Article 7 exhaustively lists six possible legal bases, including Article 7(f), the balancing provision.

In short, a data controller can rely on the balancing provision when personal data processing is necessary for the legitimate interests of the controller, or of a third party to whom the data are disclosed, unless those interests are overridden by the data subject's interests or fundamental rights. Hence, under Article 7(f), a balance must be struck between the interests of the data controller and the data subject.

The Bundesgerichtshof questions, in short, whether the balancing provision allows a website publisher to store IP addresses after a website visit.[36] Therefore, the Bundesgerichtshof's second question reads as follows:

> Does Article 7(f) of [the Data Protection Directive] preclude a provision in national law under which a service provider may collect and use a user's personal data without his consent only to the extent necessary in order to facilitate, and charge for, the specific use of the telemedium by the user concerned, and under which the purpose of ensuring the general operability of the telemedium cannot justify use of the data beyond the end of the particular use of the telemedium?[37]

---

[35] Para 55. That is the interpretation of the Telemediengesetz favoured my most academic commentators.
[36] Para 26.
[37] Para 30 amendment between square brackets by the author. See for the question as rephrased by the CJEU: para 50.

410## 5 CJEU on the balancing provision (article 7(f) of the Data Protection Directive)

The CJEU notes that a website publisher may have a legitimate interest in ensuring the continued functioning of its website, and thus in protecting the website against denial-of-service attacks.[38]

The CJEU repeats the main points from its *ASNEF* judgment (2011) on the balancing provision (Article 7(f) of the directive).[39] In *ASNEF*, the CJEU stated that the Article 7 'sets out an exhaustive and restrictive list of cases in which the processing of personal data can be regarded as being lawful.'[40] Moreover, 'Member States (…) cannot introduce principles relating to the lawfulness of the processing of personal data other than those listed in Article 7 (…), nor can they amend, by additional requirements, the scope of the six principles provided for in Article 7.'[41]

The CJEU concludes: 'Article 7(f) of [the Data protection Directive] precludes Member States from excluding, categorically and in general, the possibility of processing certain categories of personal data without allowing the opposing rights and interests at issue to be balanced against each other in a particular case. Thus, Member States cannot definitively prescribe, for certain categories of personal data, the result of the balancing of the opposing rights and interests, without allowing a different result by virtue of the particular circumstances of an individual case.'[42]

The Telemediengesetz is more restrictive than the directive's balancing provision (Article 7(f)). Therefore, the CJEU finds that the balancing provision does not allow the provision in the Telemediengesetz. The CJEU answers the German question as follows:

---

[38] Para 60.
[39] CJEU, judgment of 24 November 2011, *ASNEF and FECEMD* (C-468/10 and C-469/10, EU:C:2011:777).
[40] Para 57.
[41] Para 58. The CJEU adds that the German Federal Institutions (the website publishers) cannot, in this case, rely on the exceptions in Article 3 of the Data Protection Directive. See para 51-53.
[42] Para 62.

> Article 7(f) of [the Data Protection Directive] must be interpreted as meaning that it precludes the legislation of a Member State under which an online media services provider may collect and use personal data relating to a user of those service, without his consent, only in so far as the collection and use of that information are necessary to facilitate and charge for the specific use of those services by that user, even though the objective aiming to ensure the general operability of those services may justify the use of those data after consultation of those websites.[43]

## 6 Comment

### 6.1 Identifiability

Since the 2011 Scarlet/Sabam judgment we knew that an IP address is generally a piece of personal data, if the IP address is in the hands of the internet access provider that offers its service to the relevant individual.[44] An internet access provider assigns IP addresses to its customers, and keeps a record of this allocation.[45] The Breyer case concerns a different situation than Scarlet/Sabam: the website publisher cannot tie a name to Breyer's IP address, but Breyer's internet access provider can tie his name to the IP address.

The Breyer judgment shows that the CJEU favours an objective approach to identifiability. The CJEU's approach follows directly from recital 26, which says that 'to determine whether a person is identifiable, account should be taken of all the means likely reasonably to be used either by the controller *or by any other person to*

---

[43] Para 64 and dictum.
[44] CJEU, judgment of 24 November 2011, *Scarlet Extended* (C-70/10, EU:C:2011:7, para 51. See: S. Kulk and F.J. Zuiderveen Borgesius, 'Freedom of expression and ''right to be forgotten'' cases in the Netherlands after Google Spain', European Data Protection Law Review 2015-2, p. 113-124.
[45] Para 2 of AG opinion.

*identify the said person.*'[46] The Article 29 Working Party, in which European Data protection Authorities cooperate, also favours an objective approach to identifiability.[47]

The Breyer judgment also gives some guidance on the meaning of the phrase 'all the means likely reasonably to be used'. We now know that 'the means likely reasonably' do not include identification prohibited by law. But this threshold is satisfied if an organisation holding information about a person has legal means to obtain, from another party, extra information that enables the organization to identify the person. All in all, the CJEU's answer to the identifiability question is not surprising.

Similarly, it is not surprising that the CJEU reaffirms that Member States may not introduce laws that limit the scope of the Data Protection Directive's balancing provision (Article 7(f)).

In many situations, website publishers can rely on the balancing provision to store website visitors' IP addresses for a limited period, without the consent of the visitors. For instance, under certain circumstances, a website publisher can rely on the provision for storing IP addresses for security reasons, or for storing IP addresses to block trolls from a web forum.[48]

---

[46] The CJEU often refers to recitals to interpret the Data Protection Directive's provisions. See e.g. Case C-131/12, *Google Spain*, EU:C:2013:424, para 48, 54, 58, 66-67; Case C-101/01, *Lindqvist*, EU:C:2003:596, para 95. See generally on recitals: General Secretariat Of The Council Of The European Union, 'Manual Of Precedents For Acts Established Within The Council Of The European Union', 9 July 2010, <http://ec.europa.eu/translation/maltese/guidelines/documents/form_acts_en.pdf> accessed 24 January 2016; Tadas Klimas and Jūrateė Vaičiukaitė, 'The Law of Recitals in European Community Legislation' ILSA Journal of International & Comparative Law (2008)(15), p. 2-33.

[47] Article 29 Working Party, 'Opinion 05/2014 on Anonymisation Techniques' (WP 216) 10 April 2014, p. 9: 'it is critical to understand that when a data controller does not delete the original (identifiable) data at event-level, and the data controller hands over part of this dataset (for example after removal or masking of identifiable data), the resulting dataset is still personal data.' In a 2013 opinion the CJEU's Advocate General sees IP addresses as personal data when they're in the hands of Google. This also suggests an objective approach (Opinion AG (25 June 2013), C-131/12, Google Spain, para 48). The CJEU has neither confirmed nor disproved this view in the subsequent judgment.

[48] See in detail on Article 7(f): Article 29 Working Party, 'Opinion 06/2014 on the notion of legitimate interests of the data controller under article 7 of Directive 95/46/EC' (WP 217) 9 April 2014.

13## 6.2 Identifiability: what the Breyer case was not about

The Bundesgerichtshof asked a specific question, with a narrow scope. Below I highlight four topics that the CJEU did not explicitly discuss in Breyer.

As the Advocate General notes, the Bundesgerichtshof did not ask the following questions: '(a) whether static IP addresses are personal data under Directive 95/46;[49] (b) whether dynamic IP addresses are, always and in all circumstances, personal data within the meaning of that directive and, lastly; (c) whether the classification of dynamic IP addresses as personal data is necessary as soon as there is a third party, irrespective of who it may be, capable of using those dynamic IP addresses to identify network users.'[50] Below I also highlight (d): the CJEU did not discuss 'singling out' an individual.

(a) The Advocate General and the CJEU conclude that the Breyer case only concerns dynamic IP addresses.[51] Static IP addresses remain the same for longer periods. Hence, a website publisher could recognise returning visitors, over a longer period, on the basis of their static IP addresses. Static IP addresses thus make identifying website visitors easier, compared to dynamic IP addresses. Therefore, in general, a static IP address should sooner be regarded as a piece of personal data than a dynamic IP address.

(b) The Bundesgerichtshof did not ask whether dynamic IP addresses are personal data *in all circumstances*. The Bundesgerichtshof only asked for guidance for a situation in which (i) the website publisher does not have identifying information on the website visitor (except for his dynamic IP address), and (ii) the internet access provider has identifying information that can be tied to the IP address.

---

[49] Original footnote: "An issue addressed by the Court in the judgments in *Scarlet Extended* (C-70/10, EU:C:2011:771, paragraph 51), and *Bonnier Audio and Others* (C-461/10, EU:C:2012:219). In paragraphs 51 and 52 of the latter judgment, the Court held that communication 'of the name and address of an Internet … user using the IP address from which it is presumed that an unlawful exchange of files containing protected works took place, in order to identify that person ... constitutes the processing of personal data within the meaning of the first paragraph of Article 2 of Directive 2002/58, read in conjunction with Article 2(b) of Directive 95/46'."
[50] Para 50 of AG opinion.
[51] Para 46 of the judgment; para 47 of the AG opinion. See also AG Opinion, footnote 10.

In some situations, many website visitors might use the same IP address. For instance, at some offices, all employee computers have the same IP address.[52] In such cases, a mere IP address may not be enough to allow a website publisher to identify a visitor (from its internal access logs).

However, if such an office would be able and willing to tell a website publisher which employee visited a website, the situation would resemble the Breyer case: the employee would be identifiable. Moreover, if a website publisher uses persistent cookies, device fingerprintings, or similar unique identifiers, the publisher may be able to identify visitors, even if the visitor's IP address is not unique for that visitor.[53]

(c) The Bundesgerichtshof did not ask whether a dynamic IP address is a piece of personal data, as long as *any* party can tie a name to that IP address.

In some situations, dynamic IP addresses in the hands of a website publisher may not be personal data. Suppose a website visitor from a foreign country uses an internet access provider that refuses to cooperate with the website publisher, and does not comply with court orders. In such a situation, obtaining additional information from that access provider might not be 'a means likely reasonably to be used' by the website publisher.[54]

(d) Sometimes a company processes data to single out an individual, or to distinguish an individual within a group, but it would be difficult for anybody to tie a name to the data. For instance, many behavioural targeting companies use tracking cookies or other unique identifiers to track specific individuals, to build a profile of specific individuals, and to target advertising to specific individuals.[55]

---

[52] And all internet traffic from the country Qatar was routed through a couple of IP addresses. Zittrain JL, *The Future of the Internet and How to Stop It* (Yale University Press 2008), p. 157.
[53] See generally on tracking technologies: Mika D Ayenson, Nathaniel Good, Chris Jay Hoofnagle, Ashkan Soltani, Dietrich J. Wambach, 'Behavioral advertising: The offer you cannot refuse' (2012) *Harvard Law & Policy Review*, 6(2), 273–296; Günes Acar, Marc Juarez, Nick Nikiforakis, Claudia Diaz, Seda Gürses, Frank Piessens and Bart Preneel, 'FPDetective: Dusting the web for fingerprinters' (2013) CCS '13 Proceedings of the 2013 ACM SIGSAC conference on Computer & Communications Security 1129.
[54] See para 45 of the judgment, and recital 26 of the Data Protection Directive.

In some situations, it might be hard for anyone to add a name to such a behavioural targeting profile. But such situations are probably rare. For instance, usually behavioural targeting companies also know the IP address of the individual.[56] And the individual may have shared his or her name, email address, or other identifying information with a company – such data could be tied to the profile. For argument's sake, we assume for now that it would be impossible for anybody to tie a name to the profile.

The Article 29 Working Party takes the view that a company processes personal data if it uses data to *single out* a person, even if no name can be tied to these data.[57] Indeed, it makes sense to see singling out as identifying an individual.[58]

In Breyer, the CJEU has neither confirmed nor rejected the 'single out' approach to identifiability. The CJEU had to answer a different question. However, the Breyer judgment confirms that the CJEU favours a broad interpretation of the personal data definition. Aside from this, the General Data Protection Regulation (GDPR) mentions singling out as a way of identifying an individual, as we will see in the next section.

---

[55] See: F.J. Zuiderveen Borgesius, 'Singling out people without knowing their names – behavioural targeting, pseudonymous data, and the new Data Protection Regulation', Computer Law & Security Review, 2016-32-2, p. 256-271.
[56] The IP address is usually required to send ads to the individual's computer.
[57] The Working Party says that cookies and similar files with unique identifiers are personal data, because 'such unique identifiers enable data subjects to be "singled out" for the purpose of tracking user behaviour while browsing on different websites and thus qualify as personal data' (Article 29 Working Party, 'Opinion 16/2011 on EASA/IAB Best Practice Recommendation on Online Behavioural Advertising' (WP 188) 8 December 2011, p. 8. See also Article 29 Working Party, 'Opinion 4/2007 on the concept of personal data' (WP 136), 20 June 2007; Article 29 Working Party, 'Opinion 1/2008 on data protection issues related to search engines' (WP 148), 4 April 2008).
[58] See: F.J. Zuiderveen Borgesius, 'Singling out people without knowing their names – behavioural targeting, pseudonymous data, and the new Data Protection Regulation', Computer Law & Security Review, 2016-32-2, p. 256-271. See for another opinion: See Zwenne GJ, *De verwaterde privacywet [Diluted Privacy Law], Inaugural lecture of Prof. Dr. G. J. Zwenne to the office of Professor of Law and the Information Society at the University of Leiden on Friday, 12 April 2013* (Universiteit Leiden 2013).

**6.3    Identifiability and the GDPR**

If the CJEU had applied the (GDPR, rather than the Data Protection Directive, it would almost certainly have reached a similar conclusion. The GDPR defines personal data as follows:

> 'Personal data' means any information relating to an identified or identifiable natural person ('data subject'); an identifiable natural person is one who can be identified, directly or indirectly, in particular by reference to an identifier such as a name, an identification number, location data, an online identifier or to one or more factors specific to the physical, physiological, genetic, mental, economic, cultural or social identity of that natural person'[59]

Compared to the Data Protection Directive, the GDPR's personal data definition adds new examples of identifiers, including 'online identifier'. An IP address (and for instance a unique ID in a tracking cookie) is an online identifier. Hence, that would provide another argument to regard, in general, IP addresses as personal data.

The GDPR uses, like the CJEU and the Data Protection Directive, an objective criterion for identifiability: 'To determine whether a natural person is identifiable, account should be taken of all the means reasonably likely to be used, *such as singling out*, either by the controller *or by another person to identify the natural person directly or indirectly*.'[60] Furthermore, the GDPR states that pseudonymous

---

[59] Article 4(1) of the General Data Protection Regulation.
[60] Recital 26 of the General Data Protection Regulation.
During the drafting of the GDPR, rapporteur Jan Albrecht proposed to include the 'single out' concept in the personal data definition: "data subject' means an identified natural person or a natural person who can be identified *or singled out*, directly or indirectly, *alone or in combination with associated data*, by means reasonably likely to be used by the controller or by any other natural or legal person, in particular by reference to *a unique identifier*, location data, online identifier or to one or more factors specific to the physical, physiological, genetic, mental, economic, cultural, social *or gender* identity *or sexual orientation* of that person" (the emphasised words were proposed) (amendment 84, article 4(1), Draft Albrecht report, data can still be personal data.[61] In sum, the broad scope of the personal data definition is retained under the GDPR.

## 7  Conclusion

In summary, the main message of the Breyer case is as follows. A dynamic IP address in the hands of a website publisher is a piece of personal data for that publisher, if the publisher has the legal means enabling it to identify the visitor by obtaining additional information from that internet access provider. Two broader conclusions can also be drawn. The CJEU uses an objective criterion to determine identifiability. The CJEU favours a broad interpretation of the concept of personal data.

* * *

---

http://www.europarl.europa.eu/meetdocs/2009_2014/documents/libe/pr/922/922387/922387en.pdf accessed 24 February 2017).
[61] Article 4(5) and recital 26 of the General Data Protection Regulation.